\documentclass[aps,prl,superscriptaddress]{revtex4}
\usepackage{amsmath}
\usepackage{amssymb}
\usepackage{color}
\usepackage[dvips]{graphicx}

\begin{document}

\title{Azimuthal distinguishability of entangled photons generated in spontaneous parametric down-conversion}
\author{Clara I. Osorio}
\affiliation{ICFO-Institut de Ciencies Fotoniques, Mediterranean
Technology Park, 08860 Castelldefels (Barcelona), Spain}
\author{Gabriel Molina-Terriza}
\affiliation{ICFO-Institut de Ciencies Fotoniques, Mediterranean
Technology Park, 08860 Castelldefels (Barcelona), Spain}
\affiliation{ICREA-Institucio Catalana de Recerca i Estudis
Avancats, 08010 Barcelona, Spain}
\author{Blanca G. Font}
\affiliation{ICFO-Institut de Ciencies Fotoniques, Mediterranean
Technology Park, 08860 Castelldefels (Barcelona), Spain}
\affiliation{Dept. Signal Theory and Communications, Universitat
Politecnica de Catalunya, Jordi Girona 1-3, 08034 Barcelona Spain}
\author{Juan P. Torres}
\affiliation{ICFO-Institut de Ciencies Fotoniques, Mediterranean
Technology Park, 08860 Castelldefels (Barcelona), Spain}
\affiliation{Dept. Signal Theory and Communications, Universitat
Politecnica de Catalunya, Jordi Girona 1-3, 08034 Barcelona Spain}

\email{clara.ines.osorio@icfo.es}

\begin{abstract}
We experimentally demonstrate that paired photons generated in
different sections of a down-conversion cone, when some of the
interacting waves show Poynting vector walk-off, carry different
spatial correlations, and therefore a different degree of spatial
entanglement. This is shown to be in agreement with theoretical
results. We also discuss how this {\em azimuthal distinguishing}
information of the down-conversion cone is relevant for the
implementation of quantum sources aimed at the generation of
entanglement in other degrees of freedom, such as polarization.
 \end{abstract}

\maketitle

\section{Introduction}
Paired photons entangled in the spatial degree of freedom are
represented by an infinite dimensional Hilbert space. This offers
the possibility to implement quantum algorithms that either
inherently use dimensions higher than two or exhibit enhanced
efficiency in increasingly higher dimensions (see \cite{nature1}
and references inside). These include the demonstration of the
violation of bipartite, three dimensional Bell inequalities
\cite{vaziri1}, the implementation of the {\it quantum coin
tossing} protocol with qutrits \cite{molina1}, and the generation
of quantum states in ultra-high dimensional spaces
\cite{barreiro1}. Actually, the amount of spatial bandwidth, and
the degree of spatial entanglement, can be tailored
\cite{torres1,eberly1}, being possible to control the effective
dimensionality where spatial entanglement resides.

The most widely used source for generating paired photons with
entangled spatial properties is spontaneous parametric
down-conversion (SPDC) \cite{arnaut1,mair1}. In this process,
photons are known to be emitted in cones whose shape depends of
the phase matching conditions inside the nonlinear crystal. All
relevant experiments reported to date make use of a small section
of the full down-conversion cone. But the spatial properties of
different sections of the cone have been unexplored experimentally
up to now. This could be done, for example, by relocating the
single photon counting modules. Then, one question naturally
arises: {\em Are the entangled spatial properties of the photons
modified depending of the location in the down-conversion cone
where they are detected?}

The answer to this question is of great relevance for the
implementation of many quantum information schemes. When
considering entanglement in the spatial degree of freedom, one
should determine whether pairs of photons with different azimuthal
angle of emission might show different spatial quantum
correlations, since all quantum information applications are based
on the availability and use of specific quantum states.

Additionally, the spatial properties of entangled two-photon
states have to be taken into account even when entanglement takes
place in other degrees of freedom, such as polarization. In
general, it is required to suppress any spatial ``which-path''
information that otherwise degrades the degree of entanglement.
This is especially true for configurations that make use of a
large spatial bandwidth \cite{lee1} and in certain SPDC
configurations where horizontally and vertically polarized photons
are generated in different sections of the down-conversion cone
\cite{kwiat1, kwiat2}. Finally, the generation of heralded single
photons with well defined spatial properties, i.e. a gaussian
shape for optimum coupling into monomode optical fibers, depends
on the angle of emission \cite{torres2}.

Here we experimentally demonstrate that the presence of Poynting
vector walk-off, which is unavoidable in most SPDC configurations
currently being used, introduces {\em azimuthal distinguishing
information in the down-conversion cone}. Paired photons generated
with different azimuthal angles show correspondingly different
spatial quantum correlations and amount of entanglement. We also
show that this spatial distinguishing information can severely
degrade the quality of polarization entanglement, since the full
quantum state that describes the entangled photons is a
nonseparable mixture of polarization and spatial variables.

\begin{figure}
\centering
\includegraphics[scale=0.60]{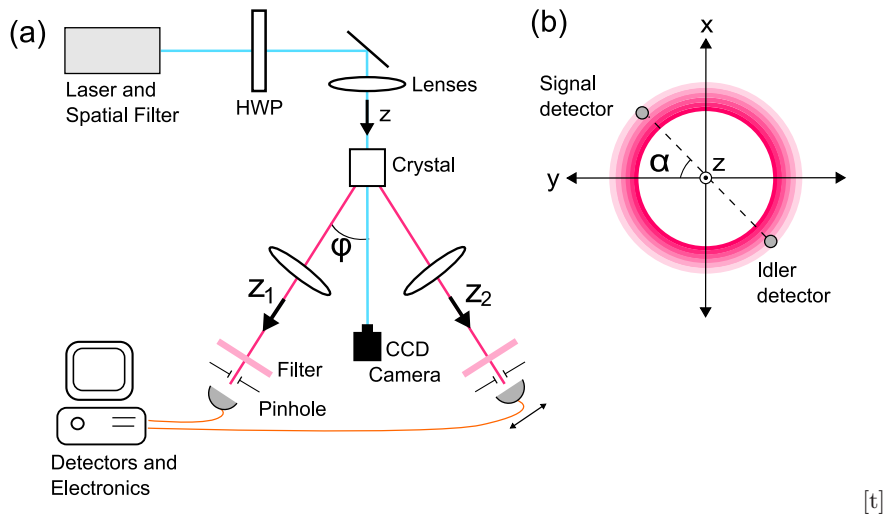}[t]
\caption{(a) Diagram of the experimental set up and (b) The
down-conversion cone. Single photon detectors are located in
opposite sides of the cone, forming an angle $\alpha$ with the
$YZ$ plane.} \label{figSPDC}
\end{figure}

\section{Experimental set-up and results}
In Fig. \ref{figSPDC} we present a scheme of our experimental
set-up. The output beam of a CW diode laser emitting at
$\lambda_p=405nm$, is appropriately spatially filtered to obtain a
beam with a gaussian profile, while a half wave plate (HWP) is
used to control the polarization. The pump beam is focalized to
$w_0=136\mu m$ beam waist on the input face of a $L=5$mm thick
lithium iodate crystal, cut at $42^{\circ}$ for Type I degenerate
collinear phase matching. The generated photons, signal and idler,
are ordinary polarized, in opposition to the extraordinary
polarized pump beam. The crystal is tilted to generate paired
photons which propagate inside the nonlinear crystal with a
non-collinear angle of $\varphi=4^{\circ}$. Due to the crystal
birefringence, the pump beam exhibits Pointing vector walk-off
with angle $\mathbf{\rho_0}=4.9^{\circ}$, while the generated
photons do not exhibit spatial walk-off. Fig. \ref{figSPDC}(b)
represents the transversal section of the down-conversion cone.
The directions of propagation of the signal and the idler photons
over this ring are determined by the azimuthal angle $\alpha$,
which is the angle between the plane of propagation of the
down-converted photons and the $YZ$ plane. To determine
experimentally the position of the crystal optics axis, and the
origin of $\alpha$, we measure the relative position of the pump
beam in the plane $XY$ at the input and output faces of the
nonlinear crystal using a CCD camera.

Right after the crystal, each of the generated photons traverse a
$2-f$ system with focal length f=$50cm$. Low-pass filters are used
to remove the remaining pump beam radiation. After the filters,
the photons are coupled into multimode fibers. In order to
increase our spatial resolution, we use small pinholes of
$300\mu$m of diameter. We keep the idler pinhole fixed and measure
the coincidence rate while scanning the signal photon transverse
spatial shape with a motorized $XY$ translation stage. Finally, as
we are interested in the different spatial correlations at
different azimuthal positions of the downconversion ring, instead
of rotating the whole detection system, the nonlinear crystal and
the polarization of the pump beam are rotated around the
propagation direction. Due to slight misalignments of the rotation
axis of the crystal, after every rotation it is necessary to
adjust the tilt of the crystal to achieve generation of photons at
the same non-collinear angle in all the cases.

\begin{figure}[ht]
\centering
\includegraphics[bb=28   331   447   581, scale=0.65]{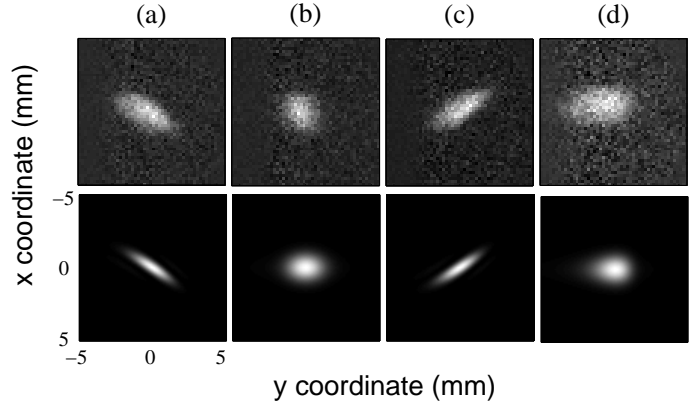}
\caption{Images showing the spatial shape of the mode function of
the signal photon when measuring coincidences rates. Upper row
corresponds to theoretical predictions, and the lower row
corresponds to experimental data. (a) $\alpha=0^{\circ}$; (b)
$\alpha=90^{\circ}$; (c) $\alpha=180^{\circ}$ and (d)
$\alpha=270^{\circ}$. See also the corresponding movie.
\label{figresults}}
\end{figure}

Images for different azimuthal sections of the cone were taken. We
present a sample of them in the upper row of Fig.
\ref{figresults}, which summarizes our main experimental results.
Each column shows the coincidence rate for $\alpha=0^{\circ}$,
$90^{\circ}$, $180^{\circ}$ and $270^{\circ}$. The movie shows the
experimental and theoretical spatial shape of the signal photon
corresponding to other values of the angle $\alpha$. Each point of
these images corresponds to the recording of a $10s$ measurement.
The typical number of coincidences at the maximum is around $10$
photons per second. The resolution of the experimental images is
$50 \times 50$ pixels. The different spatial shapes measured of
the mode function of the signal photons clearly show that the
down-conversion cone does not posses azimuthal symmetry. This
agrees with the theoretical predictions presented in the lower row
of Fig. \ref{figresults}. Note that no fitting parameter has been
used whatsoever. Slight discrepancies between experimental data
and theoretical predictions might be due to the small, but not
negligible, bandwidth of the pump beam and to the fact that the
resolution of our system is limited by the detection pinholes
size.

An interesting feature that can be observed in these images is
that the mode function in Fig. \ref{figresults}(b), corresponding
to the case $\alpha=90^{\circ}$ presents a nearly gaussian shape.
We will show below that this effect happens whenever $\varphi
\simeq \rho_0$, which corresponds to our experimental conditions.
On the other hand the mode function shown in Figs.
\ref{figresults} (a) and (c) are highly elliptical.

\section{Azimuthal distinguishability of paired photons generated in different
sections of the the down-conversion cone} To gain further insight,
we turn to the theoretical description of this problem. The signal
photon propagates along the direction ${\mathbf z_1}$ (see Fig.
\ref{figSPDC}) with longitudinal wavevector $k_s=[(\omega_s
n_s/c)^2-|\mathbf{p}|^2]^{1/2}$, and transverse wavevector ${\bf
p}=(p_x,p_y)$. Similarly, the idler photon propagates along the
${\mathbf z_2}$ direction with longitudinal wavevector $k_i$, and
transverse wavevector ${\bf q}$. Here we consider the signal and
idler photons as purely monochromatic, due to the use of a narrow
pinhole in the idler side, which selects a very small bandwidth of
frequencies of the down-converted ring. Although photons detected
in different parts of the down-conversion cone might present
slightly different polarizations \cite{migdall1}, this is a small
effect, and therefore we neglect it.

The quantum two-photon state at the output face of the nonlinear
crystal, within the first order perturbation theory, can be
written as $|\Psi\rangle=\int d {\bf p} d {\bf q} \Phi ({\bf
p},{\bf q})a_s^{\dag} ({\bf p}) a_i^{\dag} ({\bf q}) |0,0\rangle$,
where the mode function writes \cite{rubin1,torres2}
\begin{eqnarray}
& & \Phi \left( {\bf p},{\bf q} \right)={\cal N} \exp \left\{
-\frac{\left( \Gamma L \right)^2}{4} \Delta_k^2+ i \frac{\Delta_k
L}{2} \right\} \nonumber \\
& &  \times  \exp \left\{ -\frac{ \left( p_x+q_x \right)^2 w_0^2 +
\left( p_y+q_y \right)^2 w_0^2 \cos^2 \varphi}{4}  \right\}
\nonumber \\
& & \times \exp \left\{ -\frac{|{\bf p}|^2 w_s^2}{4}-\frac{|{\bf
q}|^2 w_s^2}{4} \right\}  \label{eqmo2}
\end{eqnarray}
where $\Delta_k=\tan \rho_0 \left[ (p_x+q_x)\cos \alpha +(p_y+q_y)
\cos \varphi \sin\alpha \right]-(p_y-q_y)\sin\varphi$ comes from
the phase matching condition in the $z$ direction, ${\cal N}$ is a
normalization constant, and we assume that the pump beam shows a
gaussian beam profile with beam width $w_0$ at the input face of
the nonlinear crystal. We neglect the transverse momentum
dependence of all longitudinal wavevectors. The phase matching
function, $sinc(\Delta_k L/2)$ has been approximated by an
exponential function that has the same width at the $1/e^{2}$ of
the intensity: $sinc(b x)\simeq \exp[-(\Gamma b)^{2}x^{2}]$, with
$\Gamma=0.455$. The value of $w_s$ describes the effect of the
unavoidable spatial filtering produced by the specific optical
detection system used. In our experimental set-up, the probability
to detect a signal photon at $\mathbf{x_1}$ in coincidence with an
idler photon at the fixed pinhole position $\mathbf{x_2}=0$ is
given by $R_c(\mathbf{x_1},\mathbf{x_2}=0)=|\Phi \left( 2 \pi
\mathbf{x_1}/ \left( \lambda_s f \right),\mathbf{x_2}=0 \right)
|^2$.

Eq. \ref{eqmo2} shows that the spatial mode function shape shows
ellipticity. The amount of ellipticity depends on the
non-collinear configuration \cite{molina2}, and on the azimuthal
angle of emission ($\alpha$) due to the presence of spatial walk
off. {\em The latter is the cause of the azimuthal symmetry
breaking of the down-conversion cone}. Both effects turn out to be
important when the length of the crystal $L$ is larger than the
non-collinear length $L_{nc}=w_0/\sin{\varphi}$ and the walk-off
length $L_{w}=w_0/\tan{\rho_0}$. Our experimental configuration is
fully in this regime. We should notice that in a collinear SPDC
configuration, Poynting vector walk-off also introduces
ellipticity of the mode function \cite{fedorov}.

The theory also predicts the orientation of the spatial mode
function of the signal photon, as shown in \ref{figresults}. This
is given by the slope $\tan \beta$ in the $(p_x,p_y)$ plane of the
loci of perfect phase matching transverse momentum, which writes
$\tan \beta= \left( \sin \varphi-\tan \rho_0 \cos \varphi \sin
\alpha \right)/\left(\tan \rho_0 \cos \alpha \right)$. If $\varphi
\simeq \rho_0$ and $\alpha=90^\circ$, the spatial mode function of
the signal photons shows a nearly gaussian shape, due to the
compensation of the non-collinear and walk-off effects. All these
results are in agreement with experimental data in Fig.
\ref{figresults}.

This azimuthal variation of the spatial correlations can be made
clearer if we express the mode function of the signal photon,
$\Phi_s \left( {\bf p} \right)=\Phi\left( {\bf p},{\bf q=0}
\right)$ in terms of orbital angular momentum modes. The mode
function can be described by superposition of spiral harmonics
\cite{management} $\Phi_{s} \left( \rho,\varphi \right)=\left(
2\pi \right)^{-1/2} \sum_{m} a_{m} \left( \rho \right) \exp
\left(i m \varphi \right)$, where $a_{m} \left( \rho \right)=
1/(2\pi)^{1/2} \int d\varphi \Phi_s \left(\rho,\varphi \right)
\exp \left(-i m \varphi \right)$, and $\rho$ and $\varphi$ are
cylindrical coordinates in the transverse wave-number space. The
weight of the $m$-harmonic is given by $C_{m}=\int \rho d\rho
|a_{m} \left( \rho \right)|^2$.

\begin{figure}
\centering
\includegraphics[scale=0.6]{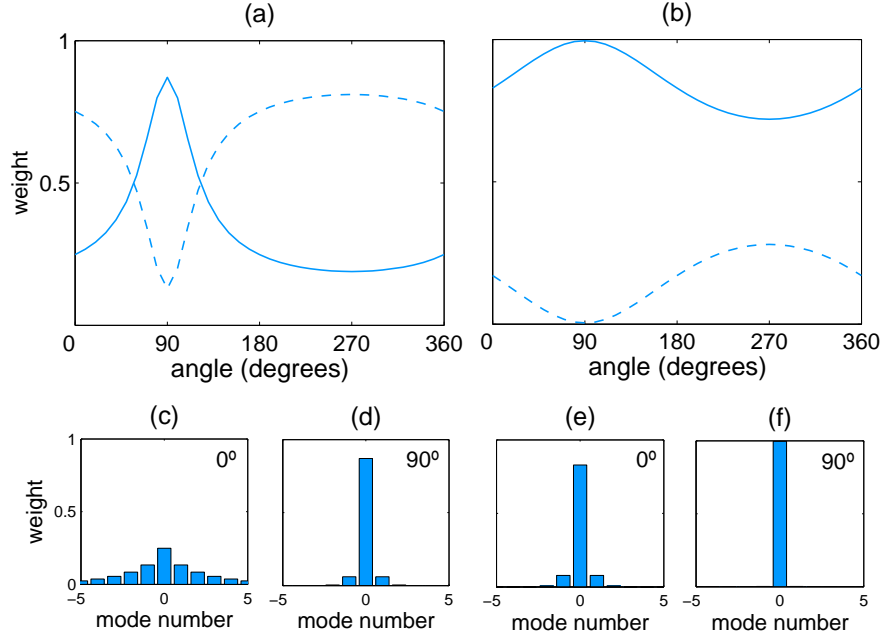}
\caption{Weight of the OAM modes $l_s=0$ (solid line), and all
other modes (dashed lines) as a function of the angle $\alpha$.
(a), (c) and (d) $w_0=100\mu m$; (b), (e) and (f) $w_0=600\mu m$.
(c) and (e) show the OAM distribution for $\alpha=0^{\circ}$, and
(d) and (f) corresponds to $\alpha=90^{\circ}$. We assume
negligible spatial filtering ($w_s \simeq 0$). Dot-dashed lines:
no spatial walk-off. \label{figmodes}}
\end{figure}

The gaussian pump beam corresponds to a mode with $l_p=0$, while
the idler photon is projected into ${\bf q}=0$, which corresponds
to projection into a large area gaussian mode ($l_i=0$). Fig.
\ref{figmodes}(a) and (b) show the weight of the mode $l_s=0$, and
the weight of all other OAM modes, as a function of the angle
$\alpha$ for two different pump beam widths. We observe that the
OAM correlations of the two-photon state change along the
down-conversion cone due to the azimuthal symmetry breaking
induced by the spatial walk-off. This implies that the
correlations between OAM modes do not follow the relationship
$l_p=l_s+l_i$. From Fig. \ref{figmodes} it is clearly observed
that for larger pump beams the azimuthal changes are smoothed out,
since in this case the non-collinear and walk-off lengths are much
larger than the crystal length.

Figure \ref{figmodes}(c) and \ref{figmodes}(d) plots the OAM
decomposition for $w_0=100 \mu$m, and Figs. \ref{figmodes}(e) and
\ref{figmodes}(f) for $w_0=600 \mu$m, for $\alpha=0,90^\circ$.
Notice that the weight of the $l_s=0$ mode is maximum for
$\alpha=90^\circ$, which therefore is the optimum angle for the
generation of heralded single photons with a gaussian-like shape.
This effect can be clearly observed in Figs. \ref{figresults}(b),
\ref{figmodes}(d) and \ref{figmodes}(f). On the contrary, for
$\alpha=270^\circ$, the combined effects of the noncollinear and
walk off effects make the weight of the $l_s=0$ mode to obtain its
minimum value. This is of relevance in any quantum information
protocol where the generated photons, no matter the degree of
freedom where the quantum information is encoded, are to be
coupled into single mode fibers.

\begin{figure}
\centering
\includegraphics[scale=0.60]{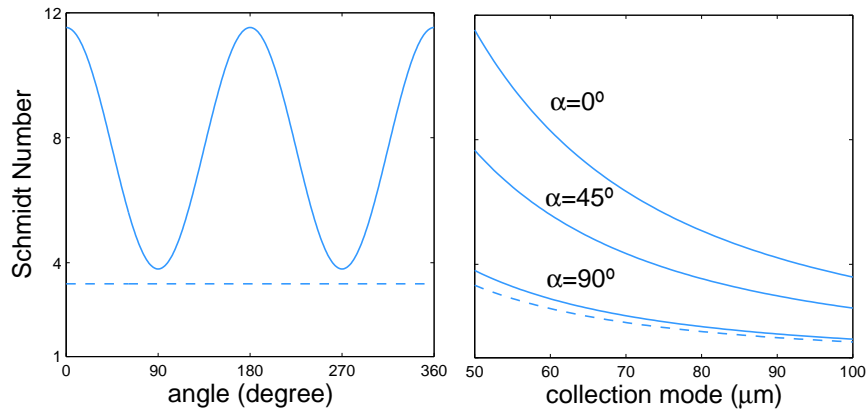}
\caption{(a) Schmidt number (K) as a function of the angle
$\alpha$. The width of the collection mode is $w_s=50 \mu$m. (b)
Schmidt number as a function of the width of the collection mode
for different values of $\alpha$. In all cases, the pump beam
width is $w_0=100 \mu$m. The Schmidt number for the case with
negligible walk-off (dashed lines) is shown for comparison.
\label{fig4}}
\end{figure}

Importantly, the degree of spatial entanglement of the two-photon
state also shows azimuthal variations, depending on the direction
of emission of the down-converted photons. Fig. \ref{fig4} shows
the Schmidt number $K=1/Tr \rho_s^2$, where $\rho_s=Tr_i
|\Psi\rangle\langle \Psi|$, is the density matrix that describe
the quantum state of the signal photon, after tracing out the
spatial variables corresponding to the idler photon. The Schmidt
number \cite{eberly1} is a measure of the degree of entanglement
of the spatial two photon state, $K=1$ corresponding to a product
state, while larger values of $K$ corresponds to increasingly
larger values of the degree of entanglement. The degree of
entanglement is maximum for $\alpha=0$, and minimum for
$\alpha=90^\circ$, as shown in Fig. \ref{fig4}(a). The degree of
entanglement is known to decrease with increasing filtering
\cite{vanexter1}, i.e., larger values of $w_s$, as shown in Fig.
\ref{fig4}(b), and to increase for larger values of the pump beam
width ($w_0$).

\section{Effects on the generation of polarization entanglement}
The azimuthal distinguishing information introduced by walking
SPDC affect the quantum properties of polarization-entangled
states, when photons generated in different sections of the
down-conversion cone are used. This is the case when using two
type I SPDC crystal whose optical axis are rotated $90^{\circ}$.
This configuration, originally demonstrated for the generation of
polarization-entangled photons \cite{kwiat1}, has been used as
well for the generation of hyperentangled quantum states
\cite{barreiro1}. The quantum state of the two-photon state writes
\begin{eqnarray}
\label{kwiat} & & |\Psi\rangle=\frac{1}{\sqrt{2}}\int d{\bf p}
d{\bf q} \left[ \Phi_1 \left({\bf p},{\bf q} \right) |H,{\bf
p}\rangle_s |H,{\bf
q}\rangle_i \nonumber \right. \\
& & \left. + \Phi_2 \left( {\bf p},{\bf q} \right)|V,{\bf
p}\rangle_s |V,{\bf q}\rangle_i \right]
\end{eqnarray}
$\Phi_1 \left({\bf p},{\bf q} \right)= \Phi \left(\alpha=0,{\bf
p},{\bf q} \right)\exp \left( i p_y \tan \rho_s L + i q_y \tan
\rho_i L\right)$ describes the spatial shape of the photons
generated in the first nonlinear crystal, $\rho_{s,i}$ are the
spatial walk-off angles of the down-converted photons traversing
the second nonlinear crystal, and $\Phi_2 \left({\bf p},{\bf q}
\right)= \Phi \left(\alpha=90^\circ,{\bf p},{\bf q} \right)$
corresponds to the photons generated in the second nonlinear
crystal. The quantum state in the polarization space is obtained
tracing out the spatial variables, i.e., $\rho_p=Tr_{s}
|\Psi\rangle \langle\Psi|$, which gives

\begin{equation}
\label{pol} \rho_p=\frac{1}{2} \left\{ |H \rangle_s |H\rangle_i
\langle H|_s \langle H |_i+ |V \rangle_s |V\rangle_i \langle V|_s
\langle V |_i+ \xi \left[ |H \rangle_s |H\rangle_i \langle V|_s
\langle V |_i+ |V \rangle_s |V\rangle_i \langle H|_s \langle H
|_i\right] \right\}
\end{equation}
where $\xi=\int d{\bf p} d{\bf q} \Phi_1 \left({\bf p},{\bf q}
\right) \Phi_2^{*} \left({\bf p},{\bf q} \right)$.

\begin{figure}
\centering
\includegraphics[scale=0.7]{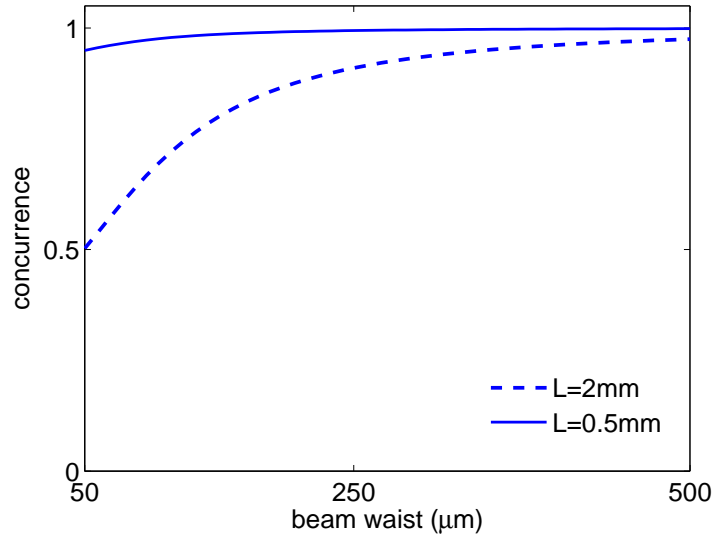}
\caption{Concurrence (C) of the polarization entangled bi-photon
state generated in a two crystal configuration, as a function of
the pump beam, for two different values of the crystal length. The
non-collinear angle is $\varphi=2^{\circ}$, and the pump beam
waist is $w_0=100 \mu$m. \label{fig5}}
\end{figure}

The degree of mixture of polarization and spatial variables is
determined by the purity ($P$) of the quantum state given by Eq.
(\ref{pol}), which writes $P=1/2 \left(1+|\xi|^2 \right)$. The
concurrence of the polarization-entangled state, which writes
writes $C=|\xi|$, quantifies the quality of the polarization
entangled state. Fig. 5 shows the concurrence of the  quantum
state for two different crystal lengths. If spatial walk-off
effects are negligible, $|\xi|=1$ and spatial and polarization
variables can be separated. Therefore, both the purity and the
concurrence are equal to $1$. This is the case shown in Fig.
\ref{fig5} for a crystal length of $L=0.5$ mm. Notwithstanding,
this is not generally he case, as demonstrated above.

Interestingly, the degree of spatial entanglement of the
horizontally polarized photons is unchanged when traversing the
second crystal, despite the fact that the down-converted photons
shows walk-off. Notwithstanding, the spatial correlations are
modified due to the presence of walk-off. It is this effect which
enhance spatial distinguishing information and thus degrades the
quality of polarization entanglement.

\section{Conclusions}
We have shown theoretically and experimentally, that the presence
of Poynting vector walk-off in SPDC configurations introduces
azimuthal distinguishing information of paired photons emitted in
different directions of propagation. The quantum correlations of
the spatial two-photon state and, consequently, the degree of
entanglement show azimuthal variations that are enhanced when
using highly focused pump beams and broadband spatial filters.
This breaking of the azimuthal symmetry of the down-conversion
cone has important consequences when designing and implementing
sources of paired photons with entangled properties.

\section{Acknowledgements}
We want to thank X. Vidal and M. Navascues for helpful
discussions. This work was supported by projects FIS2004-03556 and
Consolider-Ingenio 2010 QOIT from Spain, by the European
Commission under the Integrated Project Qubit Applications
(Contract No. 015848) and by the Generalitat de Catalunya.

\end{document}